\documentclass[preprintnumbers,prd,floatfix,nofootinbib,superscriptaddress]{revtex4}

\usepackage[a4paper,
			left=20mm,
			right=20mm,
			top=20mm,
			bottom=20mm]
			{geometry} % Use A4 paper and set margins

\usepackage[tbtags]{amsmath}  % AMS math
\usepackage{mathrsfs}
\usepackage{tabularx}
\usepackage{amssymb}          % AMS symbols
\usepackage{bm}               % bold math
\usepackage{graphicx}         % PostScript figures
\usepackage[export]{adjustbox}% to align images
\usepackage{hhline,multirow}  % for nicer tables
\usepackage{makecell}  % to have cells with more than one line (in tables)
\usepackage{dcolumn}          % Align table columns on decimal point
\usepackage{slashed}
\usepackage{datetime}
\usepackage{color}
\usepackage[dvipsnames]{xcolor}
\usepackage{slashed}
\usepackage{subfloat}

\usepackage{amscd}            % for extensible arrows (e.g., limits)
\usepackage{epsfig}
\usepackage{dcolumn}          % Align table columns on decimal point
\usepackage[dvipsnames]{xcolor}
\usepackage[normalem]{ulem}
\usepackage{mathtools}
\usepackage{esint}
\usepackage{comment}
\usepackage[utf8]{inputenc}

\RequirePackage{color}
\RequirePackage[colorlinks=true
,urlcolor=black%blue
,anchorcolor=black%blue
,citecolor=black%blue
,filecolor=black%blue
,linkcolor=black%blue
,menucolor=black%blue
,pagecolor=black%blue
,linktocpage=black%true
,pdfproducer=medialab
,pdfa=true
]{hyperref}

\newcommand{\nocontentsline}[3]{}
\newcommand{\tocless}[2]{\bgroup\let\addcontentsline=\nocontentsline#1{#2}\egroup}

\def\cc#1{\left\{#1\right\}}

\def\d#1{\mathop{{\rm d}#1}}
\def\D#1#2{\mathop{{\rm d}^#1 #2}}

\def\lag{\mathscr{L}}

\def\Ceu{C_{eu}}
\def\Ced{C_{ed}}
\def\Clqone{C_{\ell q}^{(1)}}
\def\Clqthree{C_{\ell q}^{(3)}}
\def\Clu{C_{\ell u}}
\def\Cld{C_{\ell d}}
\def\Cqe{C_{qe}}

\allowdisplaybreaks

\def\GeV{{\rm GeV}}
\def\TeV{{\rm TeV}}

\DeclareMathAlphabet{\mymathbb}{U}{BOONDOX-ds}{m}{n}

\allowdisplaybreaks[2]

\usepackage{soul}

\begin{document}

\title{SMEFT analysis with LHeC, FCC-eh, and EIC DIS pseudodata}
\normalsize{\textmd{\textit{Presented at DIS2023: XXX International Workshop on Deep-Inelastic Scattering and Related Subjects, Michigan State University, USA, 27-31 March 2023}}}

\author{Chiara Bissolotti}
\thanks{Electronic address: cbissolotti@anl.gov -- \href{https://orcid.org/0000-0003-3061-0144}{ORCID: 0000-0003-3061-0144}}
\affiliation{Argonne National Laboratory, Lemont, IL, USA}
% \affiliation{HEP Division, Argonne National Laboratory, 9700 S. Cass Avenue, Lemont, IL, 60439 USA}

\author{Radja Boughezal}
% \affiliation{HEP Division, Argonne National Laboratory, 9700 S. Cass Avenue, Lemont, IL, 60439 USA}
\affiliation{Argonne National Laboratory, Lemont, IL, USA}

\author{Kaan Simsek}
% \affiliation{HEP Division, Argonne National Laboratory, 9700 S. Cass Avenue, Lemont, IL, 60439 USA}
\affiliation{Argonne National Laboratory, Lemont, IL, USA}
\affiliation{Northwestern University, Evanston, IL, USA}

\begin{abstract}
In this study, we examine the possibilities opened by upcoming high-energy deep-inelastic scattering (DIS) experiments to investigate new physics within the framework of the Standard Model Effective Field Theory (SMEFT). Specifically, we investigate the beyond-the-Standard-Model (BSM) potential of the Large Hadron-electron Collider (LHeC) and the Future Circular lepton-hadron Collider (FCC-eh), and we improve previous simulations of the Electron-Ion Collider (EIC) by incorporating $Z$-boson vertex corrections. 
% The accuracy of vertex corrections at the $Z$-pole is affected by various degeneracies in the parameter space of Wilson coefficients. 
Our fits, performed using DIS pseudodata, reveal that the LHeC and the FCC-eh can play a crucial role in resolving degeneracies observed in the parameter space of Wilson coefficients in global fits using the Higgs, diboson, electroweak, and top data. This emphasizes the significance of precision DIS measurements in advancing our understanding of new physics.
\end{abstract}

\maketitle
\vspace{-0.5cm}
\section{Introduction}
\label{s:intro}

The Standard Model (SM) stands as an elegant and comprehensive theory, regarded as the most complete framework to date. However, this theory falls short of providing a complete understanding of the fundamental workings of the universe, and numerous compelling indications suggest the existence of physics beyond the Standard Model. 

Given the fact that no particle outside the SM has been found yet, employing an effective field theory (EFT) seems particularly advantageous for the exploration of BSM physics. Specifically, the Standard Model Effective Field Theory (SMEFT) emerges as a highly suitable and versatile approach, offering a model-independent framework for conducting such investigations.

In the SMEFT, higher-dimensional operators in mass dimensions are constructed utilizing the current particle spectrum of the SM. The SMEFT framework assumes that any new physics lies beyond the energy range of both the SM particles and the colliders' capabilities. A review of the SMEFT can be found in Ref.~\cite{Brivio:2017vri}.

This contribution is a summary of the results discussed in Ref.~\cite{Bissolotti:2023vdw}.
Our goal in this work is to study the BSM potential of future colliders, such as the LHeC, FCC-eh, and the EIC with a detailed accounting of anticipated uncertainties.
Following previous studies~\cite{Britzger:2020kgg}, we consider the neutral-current (NC) deep-inelastic scattering (DIS) cross section as our observable at the LHeC and at the FCC-eh, while, at the EIC, we focus on parity-violating (PV) asymmetries, as done in Ref.~\cite{Boughezal:2022pmb}.

% The leading order basis of the SMEFT for on-shell fields has been completely classified up to dimension-12~\cite{Harlander:2023psl}. In this work, we consider only dimension-6 operators.
% (there is a lepton-number violating operator at dimension-5, which is irrelevant to our study). 

Prior studies have demonstrated that DIS measurements at the EIC and in low-energy fixed target experiments have the potential to address blind sports observed in the semi-leptonic four fermion Wilson coefficient space that persist after Drell-Yan measurements at the LHC~\cite{Boughezal:2020uwq, Boughezal:2021kla}. Additionally, EIC measurements of single-spin asymmetries offer a competitive tool for probing the Wilson coefficients associated with dipole operators~\cite{Boughezal:2023ooo}.

We consider here the full spectrum of Wilson coefficients that can alter the DIS process at leading order in the SMEFT loop expansion. These include both semi-leptonic four-fermion Wilson coefficients and $Z$-boson vertex correction, for a total of 17 Wilson coefficients.

% Traditional measurements of vertex corrections have relied on Z-pole precision electroweak observables (i.e. at LEP and SLC). However, the kinematic information provided by Z-pole data is limited, resulting in multiple degeneracies among the Wilson coefficients. 
% This issue was demonstrated in a previous study (Ref.~\cite{}), which also incorporated contributions from existing LHC data. 
% Moreover, the constraints on the Wilson coefficients weaken significantly when the full spectrum of coefficients is considered, as opposed to activating only a single coefficient. 
In this study, we present evidence that future DIS measurements can play a vital role in resolving degeneracies observed in the parameter space of Wilson coefficients in global fits using the Higgs, diboson, electroweak, and top data.
 
\section{Formalism}
\label{s:form}

\subsection{SMEFT formalism}
\label{ss:SMEFTform}
The SMEFT serves as a gauge symmetry-preserving, model-independent extension of the SM Lagrangian. Within this framework, one constructs higher-dimensional operators, denoted as $O_k^{(n)}$, utilizing the existing SM particle spectrum. The associated Wilson coefficients, represented as $C_k^{(n)}$, quantify the strength of these operators. These effective couplings are defined at an ultraviolet (UV) cut-off scale, $\Lambda$, which is assumed to exceed the masses of all SM particles, as well as the energies accessible by collider experiments.
The Lagrangian takes the form
\begin{equation}
\label{e:SMEFT_lagrangian}
    \lag_{\rm SMEFT} = 
        \lag_{\rm SM}
        + \sum_{n > 4} {1 \over \Lambda^{n-4}} \sum_k C_k^{(n)} O_k^{(n)} \, .
\end{equation}
In this study, we focus exclusively on dimension-6 operators, disregarding any dimension-5 operators that violate lepton-number conservation, as they are not relevant to our analysis. %of observables that preserve lepton number. 
Our treatment of observables linearizes their dependence on the Wilson coefficients. 
There are a total of 17 operators that impact NC DIS matrix elements when considering leading-order coupling constants~\cite{Grzadkowski:2010es}, and these operators are summarized in Table \ref{tab:ops}.
\begin{table}
    \centering
    \begin{tabular}{|c|c|c|c|}
        \hline 
        \multicolumn{2}{|c|}{$ffV$} & \multicolumn{2}{|c|}{semi-leptonic four-fermion} \\
        \hline 
        $C_{\varphi WB}$ & $O_{\varphi WB} = (\varphi^\dagger \tau^I \varphi) W_{\mu\nu}^I B^{\mu\nu}$ & $\Clqone$ & $O_{\ell q}^{(1)} = (\bar \ell \gamma_\mu \ell) (\bar q \gamma^\mu q)$ \\ 
        \hline 
        $C_{\varphi D}$ & $O_{\varphi D} = (\varphi^\dagger D_\mu \varphi)^* (\varphi^\dagger D^\mu \varphi)$ & $\Clqthree$ & $(\bar \ell \gamma_\mu \tau^I \ell) (\bar q \gamma^\mu \tau^I q)$ \\
        \hline 
        $C_{\varphi \ell}^{(1)}$ & $O_{\varphi \ell}^{(1)} = (\varphi^\dagger i \stackrel{\leftrightarrow}{D}_\mu \varphi) (\bar \ell \gamma^\mu \ell)$ & $\Ceu$ & $O_{eu} = (\bar e \gamma_\mu e) (\bar u \gamma^\mu u)$ \\
        \hline 
        $C_{\varphi \ell}^{(3)}$ & $O_{\varphi \ell}^{(3)} = (\varphi ^\dagger i \stackrel{\leftrightarrow}{D}_\mu \tau^I \varphi) (\bar \ell \gamma^\mu \tau^I \ell)$ & $\Ced$ & $O_{ed} = (\bar e \gamma_\mu e) (\bar d \gamma^\mu d)$ \\ 
        \hline 
        $C_{\varphi e}$ & $O_{\varphi e} = (\varphi^\dagger i \stackrel{\leftrightarrow}{D}_\mu \varphi) (\bar e \gamma^\mu e)$ & $\Clu$ & $O_{\ell u} = (\bar \ell \gamma_\mu \ell) (\bar u \gamma^\mu u)$ \\
        \hline 
        $C_{\varphi q}^{(1)}$ & $O_{\varphi q}^{(1)} = (\varphi^\dagger i \stackrel{\leftrightarrow}{D}_\mu \varphi) (\bar q \gamma^\mu q)$ & $\Cld$ & $O_{\ell d} = (\bar \ell \gamma_\mu \ell) (\bar d \gamma^\mu d)$\\ 
        \hline 
        $C_{\varphi q}^{(3)}$ & $O_{\varphi q}^{(3)} = (\varphi ^\dagger i \stackrel{\leftrightarrow}{D}_\mu \tau^I \varphi) (\bar q \gamma^\mu \tau^I q)$ & $\Cqe$ & $O_{qe} = (\bar q \gamma_\mu q) (\bar e \gamma^\mu e)$ \\ 
        \hline 
        $C_{\varphi u}$ & $O_{\varphi u} = (\varphi^\dagger i \stackrel{\leftrightarrow}{D}_\mu \varphi) (\bar u \gamma^\mu u)$ \\ 
        \cline{1-2}
        $C_{\varphi d}$ & $O_{\varphi d} = (\varphi^\dagger i \stackrel{\leftrightarrow}{D}_\mu \varphi) (\bar d \gamma^\mu d)$  \\
        \cline{1-2}
        $C_{\ell \ell}$ & $O_{\ell \ell} = (\bar \ell \gamma_\mu \ell) (\bar \ell \gamma^\mu \ell)$  \\
        \cline{1-2} 
        \cline{1-2}
    \end{tabular} 
        \caption{Dimension-6 operators in the Warsaw basis \cite{Grzadkowski:2010es} affecting NC DIS matrix elements at leading order in the coupling constants. Operators in the left column shift the $ffV$ vertices, while those on the right induce semi-leptonic four-fermion contact interactions. Both the operators and their associated Wilson coefficients are shown. Here, $\varphi$ represents the Higgs doublet belonging to the SU(2) gauge group; $\ell$ and $q$ refer to the left-handed lepton and quark doublets, while $e$, $u$, and $d$ denote the right-handed electron, up-quark, and down-quark singlets, respectively. The notation $\tau^I$ represents the Pauli matrices; the double-arrow covariant derivative is defined as in~\cite{Bissolotti:2023vdw}.} 
        \label{tab:ops}
\end{table}
In our study, we suppress flavor indices and assume flavor universality. 
% We remark that operators containing scalar and dipole fermionic bilinears are discarded in our analysis. Such vertex factors produce cross section contributions proportional to fermion masses, which are small and are neglected here. 
We note that SMEFT one-loop corrections are anticipated to be of lesser significance compared to next-to-leading order (NLO) QCD corrections. In our analysis, we incorporate the NLO QCD corrections and observe that they have minimal impact on the obtained outcomes. Consequently, we make the assumption that the higher-order terms within the SMEFT loop expansion can be safely disregarded.

%%%%%%%%%%%%%%%%%%%%%%%%%%%%%%%%%%%%%%%%%%%%%%%%%
\subsection{DIS formalism}
\label{ss:DIS_form}

In NC DIS, a lepton scatters off a nucleon, namely $\ell + H \to \ell' + X$, where $\ell$ is an electron or a positron, $H$ can be a proton or a deuteron, and $\ell'$ and $X$ are the final-state lepton and hadronic systems, respectively. The process is mediated by a photon or a $Z$-boson exchange.
In this study, we deal with reduced cross sections, defined as 
\begin{align}
    {\D2{\sigma_{r, {\rm NC}}^\ell} \over \d x \d {Q^2}} &= 
        \cc{
            {2 \pi \alpha^2 \over x Q^4} [1 + (1-y)^2]
        }^{-1} {\D2{\sigma^\ell_{\rm NC}} \over \d x \d {Q^2}}\, , \\ 
    {\D2{\Delta \sigma_{r, {\rm NC}}^\ell} \over \d x \d {Q^2}} &= 
        \cc{
            {4 \pi \alpha^2 \over x Q^4} [1 + (1-y)^2]
        }^{-1} {\D2{\Delta\sigma^\ell_{\rm NC}} \over \d x \d {Q^2}}\, ,
\end{align}
where $Q$ is the usual DIS momentum transfer, $x$ is the Bjorken variable, and $y$ is the inelasticity parameter. The expressions for the NC DIS cross sections for collisions of a lepton $\ell$ with an unpolarized or polarized hadron, ${\D2{\sigma^\ell_{\rm NC}} \over \d x \d {Q^2}}$ and ${\D2{\Delta \sigma^\ell_{\rm NC}} \over \d x \d {Q^2}}$, are given, for example, in Ref.~\cite{Bissolotti:2023vdw}.
From this point forward, whenever we refer to cross sections, we will refer to the reduced ones and we will indicate them with $(\Delta)\sigma_{\rm NC}$.

We include NLO QCD corrections to both the SM and the SMEFT corrections. 
The NLO QCD corrections to the SM process are well known \cite{deFlorian:2012wk, Altarelli:1979kv, Vogelsang:1990ug, Altarelli:1979ub, deFlorian:1994wp}. These corrections modify only the quark lines, and, therefore, the corrections are identical for both SM and SMEFT cross sections.

%%%%%%%%%%%%%%%%%%%%%%%%%%%%%%%%%%%%%%%%%%%%%%%%%

The observable of interest at the LHeC and FCC-eh is the NC DIS cross section, $\sigma_{\rm NC}$, of unpolarized protons with electrons or positrons of various polarizations. 
For the EIC, we consider PV asymmetries in cross sections of polarized electrons with either polarized or unpolarized protons/deuterons. We define the unpolarized PV asymmetry, $A_{\rm PV}$, and the polarized one, $\Delta A_{\rm PV}$, by
\begin{align}
\label{eq:cs}
    A_{\rm PV} =
        {
        \sigma_{\rm NC}^+ - \sigma_{\rm NC}^- \over 
        \sigma_{\rm NC}^+ + \sigma_{\rm NC}^- 
        } \quad \quad \quad \quad \quad \quad
    \Delta A_{\rm PV} =
    {  
       \frac{\Delta \sigma_{\rm NC}^0}{\sigma_{\rm NC}^0} \, .
    }    
\end{align}
In Eq.(\ref{eq:cs}), $\sigma_{\rm NC}^\pm$ is the unpolarized NC DIS $e^-H$ ($H=p,D$) cross section evaluated with $\lambda_\ell = \pm P_\ell$, $\sigma_{\rm NC}^0$ is the same as $\sigma_{\rm NC}^\pm$ but with $\lambda_\ell = 0$, and $\Delta \sigma_{\rm NC}^0$ is the same as $\sigma_{\rm NC}^0$ but with a polarized hadron. $P_\ell$ is the assumed value for the lepton beam polarization at the EIC. 

In this study, we linearize the SMEFT expressions. Thus, the SMEFT observables have the generic form
\begin{align}
    \mathcal O = \mathcal O^{\rm SM} + \sum_k C_k \ \delta \mathcal O_k + \mathcal O (C_k^2)\, ,
\end{align}
where $k$ runs over the active Wilson coefficients, $\mathcal O$ is the observable, and $\delta \mathcal O_k$ is the SMEFT correction to the observable proportional to the Wilson coefficient $C_k$. 

%%%%%%%%%%%%%%%%%%%%%%%%%%%%%%%%%%%%%%%%%%%%%%%%%%%%%%%%%%%%%%%%%
\section{Pseudodata sets}
\label{s:data}
%%%%%%%%%%%%%%%%%%%%%%%%%%%%%%%%%%%%%%%%%%%%%%%%%%%%%%%%%%%%%%%%%

For our analysis, we utilize the most recent publicly available LHeC pseudodata sets~\cite{r:KleinData, r:KleinPaper}, as well as the EIC dataset that has been identified as the most sensitive to SMEFT Wilson coefficients in~\cite{Boughezal:2022pmb}.
Regarding the FCC-eh, we generate pseudodata sets using the procedure outlined in~\cite{Boughezal:2022pmb}, taking into account the FCC-eh run parameters as specified in~\cite{Britzger:2022abi}.
From this point onward, we refer to these pseudodata sets as \textit{data sets}. For a full list of the data sets included in this analysis, we point the reader to Table II of Ref.~\cite{Bissolotti:2023vdw}.
In order to minimize significant uncertainty from non-perturbative QCD and nuclear dynamics that occur at low $Q$ and high $x$, where we expect SMEFT effects to be diminished, we limit ourselves to the bins that fulfill $x\leq0.5$, $Q\geq10~\GeV$, and $0.1\leq y\leq 0.9$.

Regarding the uncertainties, we adopt the error estimates from prior assessments~\cite{Britzger:2020kgg,Britzger:2022abi} for the LHeC and the FCC-eh. 
We introduce the systematics in a completely correlated manner and consider the luminosity error to be 1\% relative to the cross section.

As for the EIC asymmetries, we take into account both statistical and systematic uncertainties. The systematic errors due to particle background and other imperfections in measurements are treated as uncorrelated and are 1\% relative to the asymmetry. We assume uncertainties in lepton (hadron) beam polarization to be fully correlated and 1\% (2\%) relative in asymmetry. More discussion on the anticipated experimental uncertainties at the EIC is given in Ref.~\cite{Boughezal:2022pmb}.

Additionally, for all the data sets, we take into account PDF errors fully correlated between bins.
In the Appendices of Ref.~\cite{Bissolotti:2023vdw}, we discuss how these systematic uncertainties are incorporated into the error matrix for our analysis; we also give details of our pseudodata generation and describe our statistical procedure for deriving the bounds on Wilson coefficients.
For a full breakdown of the uncertainties of all the data sets included in this study, we the reader may refer to Ref.~\cite{Bissolotti:2023vdw}.

\section{Results}
\label{s:results}
\vspace{-0.2cm}
\subsection{Bounds on semi-leptonic four-fermion operators}
\label{ss:4fermions}

At first, we activate solely the seven semi-leptonic four-fermion operators. Previous investigations have indicated that the Drell-Yan process at the LHC, which is naturally suited to probe these operators given its energy coverage and exceptional measurement precision, encounters challenges in disentangling specific linear combinations of Wilson coefficients within this subspace~\cite{Alte:2018xgc,Boughezal:2020uwq}. Future DIS experiments offer the potential to resolve these degeneracies and provide valuable insights in this regard~\cite{Boughezal:2020uwq,Boughezal:2022pmb}.
% Restricting ourselves to this subspace of Wilson coefficients allows us to compare the potential of DIS measurements at the EIC, FCC-eh and LHeC to improve upon Drell-Yan measurements at the LHC.
In Table~\ref{tab:1d-bounds-uv-table}, we show the marginalized and non-marginalized\footnote{For details about the marginalization process, we refer the reader to Ref.~\cite{Bissolotti:2023vdw}.} 95\% confidence-level (CL) bounds on the semi-leptonic four-fermion Wilson coefficients and the corresponding effective UV scales, expressed in TeV, obtained from the full seven-parameter ($7d$) joint fits for the EIC, LHeC, and FCC-eh data sets, respectively.

\renewcommand{\arraystretch}{1.1} % Adjust the padding around the text
\begin{table}
    \centering
    \scriptsize
    \setlength\extrarowheight{2pt} % for a more open 'look'
    \begin{tabularx}{\textwidth}{|X|*{15}{>{\centering\arraybackslash}X|}}
        \hline
        \multicolumn{2}{|c|}{} 
        & \multicolumn{2}{c|}{$\Ceu$} & \multicolumn{2}{c|}{$\Ced$} 
        & \multicolumn{2}{c|}{$\Clqone$} & \multicolumn{2}{c|}{$\Clqthree$ } 
        & \multicolumn{2}{c|}{$\Clu$} & \multicolumn{2}{c|}{$\Cld$}
        & \multicolumn{2}{c|}{$\Cqe$} \\
        \hline
        \multicolumn{2}{|c|}{} & 95\% cl & UV sc. & 95\% cl & UV sc. & 95\% cl & UV sc. & 95\% cl & UV sc. & 95\% cl & UV sc. & 95\% cl & UV sc. & 95\% cl & UV sc. \\
        \hline
        \hline
        \multirow{2}{5mm}{EIC}
            &    mar.
                & 2.1  & 0.69 & 7.2 & 0.37 & 2.8 & 0.59 & 4.2 & 0.49 & 9.1 & 0.33 & 9.8 & 0.32 & 8.9 & 0.33 \\                    
          \cline{2-16}
            &   nonmar. 
                & 0.12 & 2.9  & 0.34 & 1.7 & 0.17 & 2.4 & 0.10 & 3.2 & 0.28 & 1.9 & 0.57 & 1.3 & 0.39 & 1.6 \\  
        \hline
        \hline
        \multirow{2}{5mm}{LHeC}
            &   mar.
                & 0.0053  &  14. & 0.026 & 6.2 & 0.020 & 7.1 & 0.011 & 9.5 & 0.032 & 5.6 & 0.16 & 2.5 & 0.018 & 7.4\\                    
            \cline{2-16}
            &   nonmar. 
                & 0.0022 & 21. & 0.0097 & 10. & 0.0031 & 18. & 0.0017 & 24. & 0.0084 & 11. & 0.036 & 5.3 & 0.011 & 9.7\\  
        \hline
        \hline
        \multirow{2}{5mm}{FCC-eh}
            &   mar.
                & 0.0031 & 18. & 0.0070 & 12. & 0.035 & 5.4 & 0.014 & 8.4 & 0.068 & 3.8 & 0.26 & 2. & 0.0092 & 10.\\                   
          \cline{2-16}
            &   nonmar. 
                & 0.00056 & 42. & 0.0012 & 28. & 0.0014 & 27. & 0.00038 & 51. & 0.0028 & 19. & 0.0061 & 13. & 0.0016 & 25.\\                    
        \hline
    \end{tabularx}
    \caption{Marginalized and non-marginalized 95\% CL bounds on semi-leptonic four-fermion Wilson coefficients at $\Lambda = 1~\TeV$ for the combined EIC, LHeC, and FCC-eh datasets, as well as the corresponding effective UV scales, in units of TeV.}
    \label{tab:1d-bounds-uv-table}
\end{table}

A comprehensive table, including the results for individual data sets from each collider, can be found in Ref.~\cite{Bissolotti:2023vdw}. Examining Table~\ref{tab:1d-bounds-uv-table}, we observe that the effective scales probed in the fully marginalized joint fits vary depending on the Wilson coefficient. At the EIC, the probed scales range from 300 GeV to 700 GeV. For the LHeC, the range extends from 2.5 TeV to 14 TeV, while at the FCC-eh, the probed scales span from 2.0 TeV to 18 TeV.

Furthermore, we note that the joint LHeC data set imposes significantly stronger bounds on semi-leptonic four-fermion Wilson coefficients compared to the EIC. This difference arises due to the LHeC's higher momentum transfers, and thus deviations induced by SMEFT are more pronounced. For the majority of operators, the joint FCC-eh fit imposes stronger constraints than the joint LHeC fit.

The effective UV scales presented in Table~\ref{tab:1d-bounds-uv-table} are defined as $\Lambda/\sqrt{C_k}$ for each Wilson coefficient $C_k$. We emphasize that the convergence of the EFT expansion is governed by the ratio $C_k Q^2/\Lambda^2$, where $Q$ denotes the DIS momentum transfer. The obtained constraints on the effective scales indicate that this ratio remains significantly below unity for all considered runs. This supports our decision to truncate the expansion at dimension-6 and to linearize the dimension-6 SMEFT effects.

\begin{figure}
    [htbp]
    \centering
    \includegraphics[height=.3\textheight]{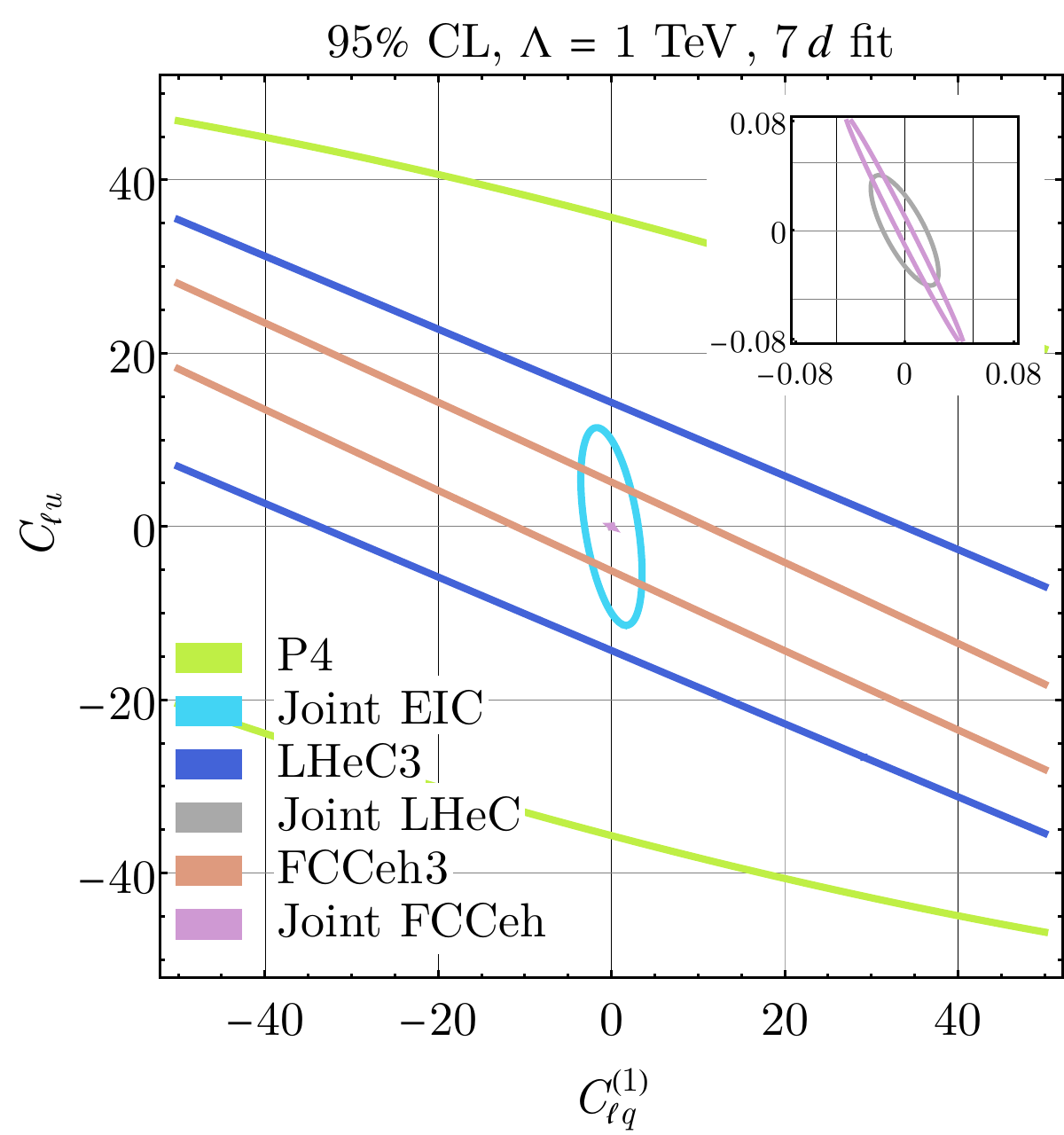}
    \includegraphics[height=.3\textheight]{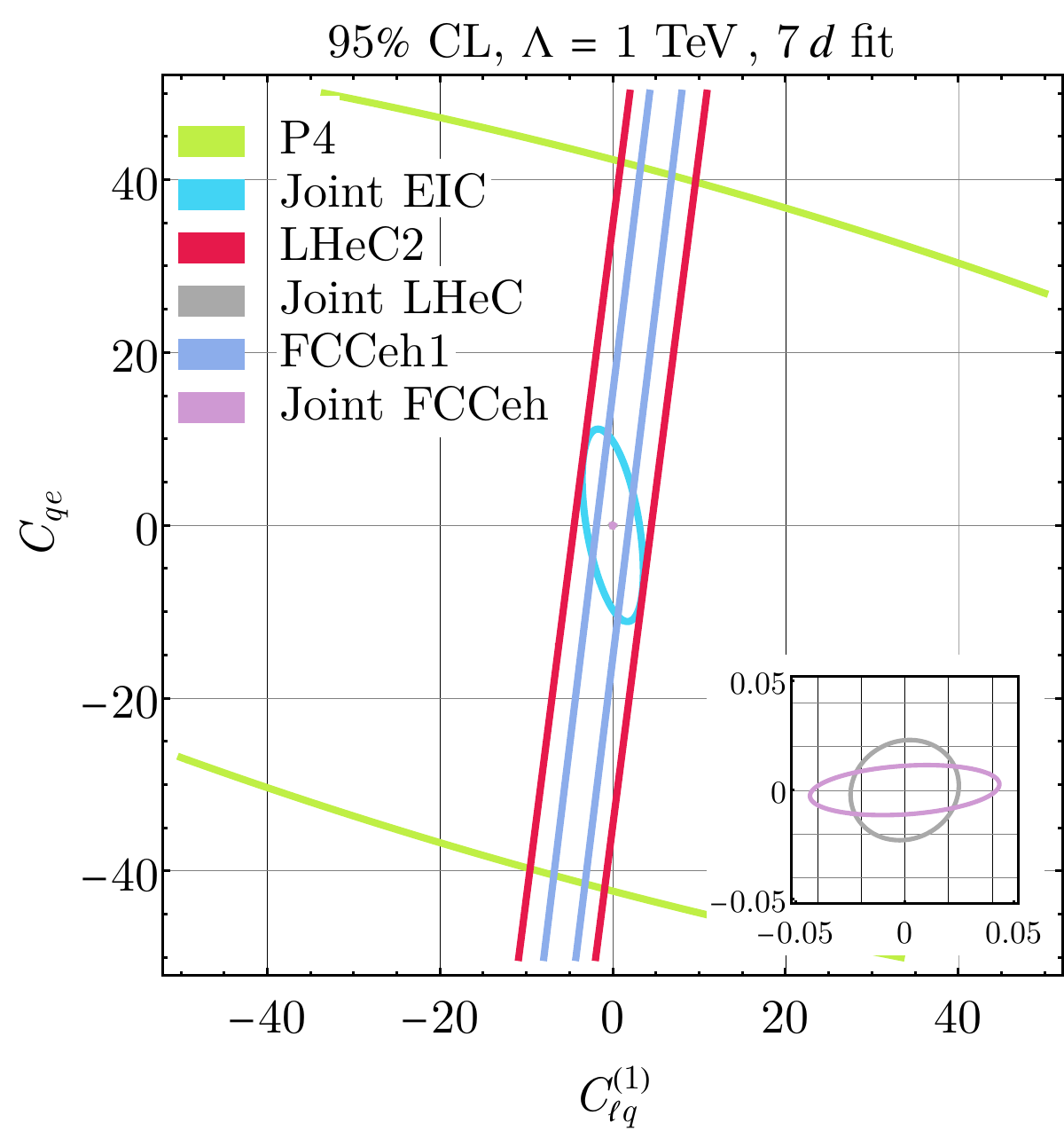}
    \caption{Marginalized 95\% CL ellipses for the parameter subspaces spanned by $\Clqone$ and $\Clu$ (left) and $\Clqone$ and $\Cqe$ (right) with $\Lambda = 1~\TeV$. In the plots, we show the strongest individual EIC data set, the strongest LHeC and FCC-eh sets for these Wilson coefficients, as well as the joint EIC, LHeC, and FCC-eh fits. The insets show the zoomed-in plots of the joint LHeC and FCC-eh fits.}
    \label{fig:clq1-clu}
\end{figure}

In Fig.~\ref{fig:clq1-clu}, we show representative confidence ellipses projected from the $7d$ fit of the four-fermion Wilson coefficients. 
We can see the emergence of flat directions for individual sets, namely LHeC3 and FCCeh3. These flat directions appear to be resolved in the joint fits.
We note that the EIC confidence ellipse remains weaker in the joint fit with respect to the LHeC and the FCC-eh.

%%%%%%%%%%%%%%%%%%%%%%%%%%%%%%%%%%%%%%%%%%%%%%%
\vspace{-0.2cm}
\subsection{Bounds on $ffV$ vertex corrections}
\label{ss:ffVcorrections}

We proceed by activating all 17 Wilson coefficients listed in Table \ref{tab:ops}, which encompass both four-fermion interactions and operators affecting the $ffV$ vertices. Generally, corrections to the $ffV$ vertices are expected to be tightly constrained by precision $Z$-pole observables. Notably, fits considering only a single activated Wilson coefficient yield remarkably stringent bounds, reaching up to 10 TeV in certain cases~\cite{Dawson:2019clf}. However, the limited number of available measurements gives rise to multiple degeneracies within this parameter space. This phenomenon is highlighted in Ref.\cite{Ellis:2020unq}, where the bounds on $ffV$ vertex corrections are significantly relaxed by approximately one order of magnitude when transitioning from single-coefficient fits to results where the remaining Wilson coefficients are marginalized over. For instance, the reach of the effective UV scale associated with the coefficient $C_{\phi WB}$ diminishes from approximately 15 TeV to 1 TeV when all coefficients are active (see Ref.\cite{Ellis:2020unq}, Fig.~3).
% Other possibilities for probing these couplings include top, Higgs and diboson data at the LHC, which are also considered in~\cite{Ellis:2020unq}, and on-shell $Z$-boson production at the LHC~\cite{Breso-Pla:2021qoe}. 
We consider here the potential of future DIS experiments to probe this sector of the SMEFT.

A table presenting the marginalized 95\% CL bounds on Wilson coefficients obtained from the full $17d$ fit can be found in Ref.~\cite{Bissolotti:2023vdw}. In that reference, we provide analogous results for the joint EIC fit, as well as the joint LHeC and FCC-eh constraints. 

We present here a summary of the main findings of the $17d$ fit. The obtained bounds from the LHeC surpass those obtained from the joint fit of electroweak precision data and the LHC results for the majority of Wilson coefficients, indicating that the inclusion of LHeC data enhances the constraining power in the global fit. Furthermore, the FCC-eh bounds are generally stronger than both the LHeC and EIC bounds in most cases.
The bounds from the EIC are weaker than those obtained from the LHeC and in~\cite{Ellis:2020unq}. It is important to note that a direct comparison between the fits conducted in~\cite{Ellis:2020unq} and our study may not be entirely straightforward due to differences in the number of fitted parameters.

\begin{figure}
    [htbp]
    \centering
    \includegraphics[height=.3\textheight]{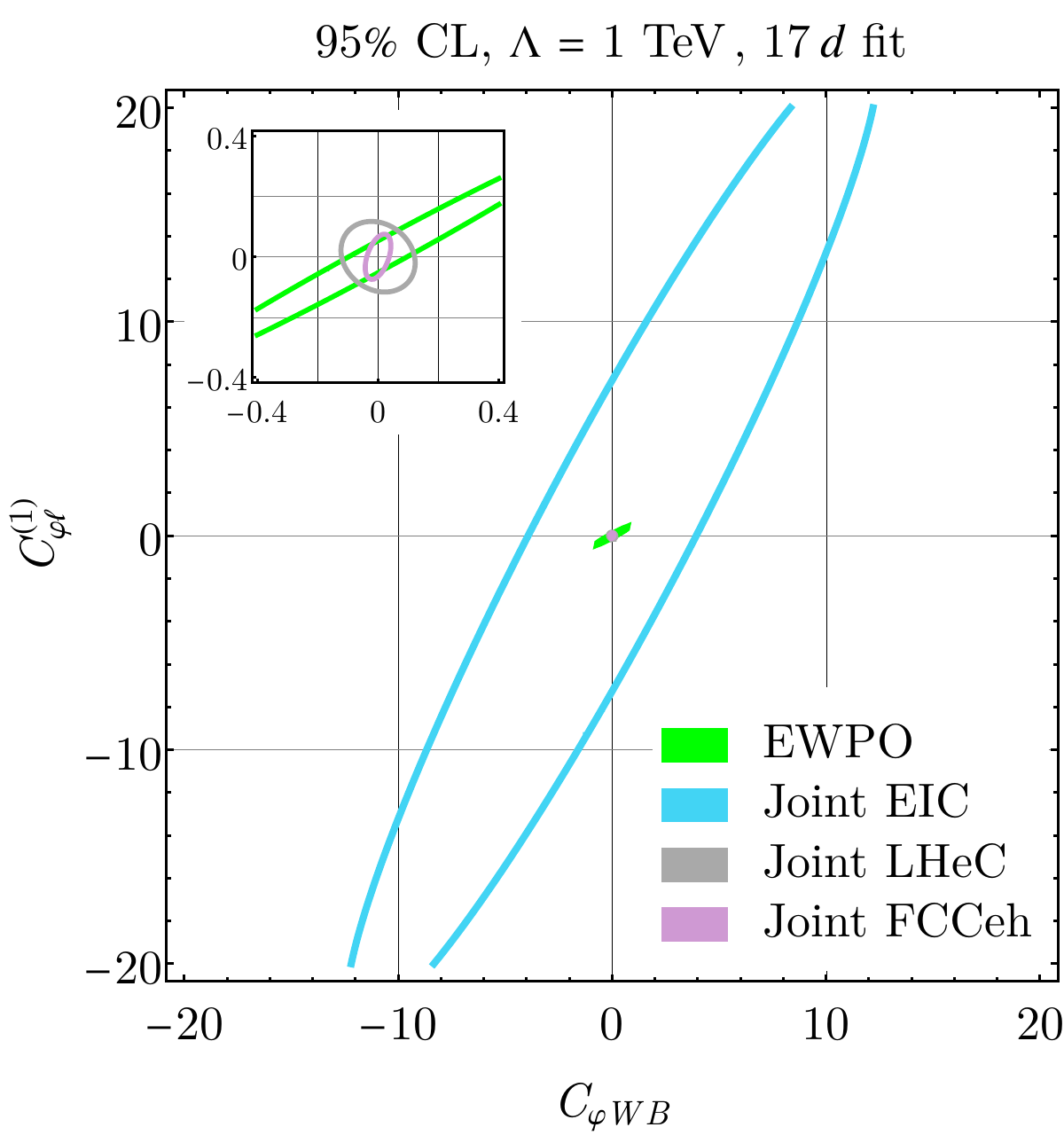}
    \includegraphics[height=.3\textheight]{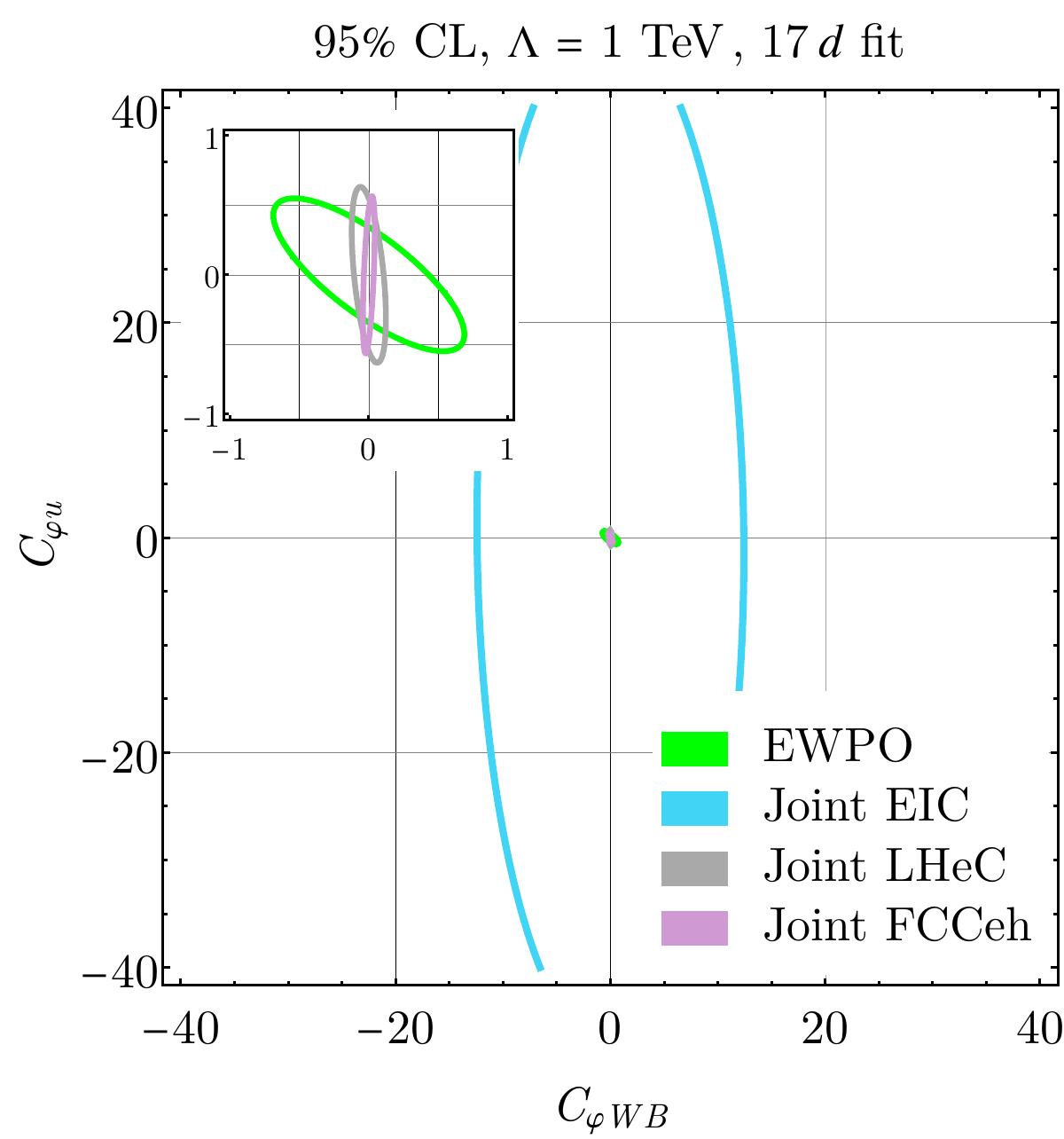}
    \caption{Marginalized 95\% CL ellipses in the two-parameter fits of $C_{\varphi WB}$ and $C_{\varphi \ell}^{(1)}$ (left) and $C_{\varphi WB}$ and $C_{\varphi u}$ (right) at $\Lambda = 1~\TeV$. Shown are the joint EIC, LHeC, and FCC-eh fits, as well as the EWPO fit adapted from Ref.~\cite{Ellis:2020unq}.}
    \label{fig:ewpo-comparison2}
\end{figure}

To further investigate the implications of including future precision DIS data in the existing global fit, we examine representative $2d$ projections of our results. In Fig.~\ref{fig:ewpo-comparison2}, we present non-marginalized 95\% CL ellipses in the parameter subspace defined by $(C_{\varphi WB}, C_{\varphi \ell}^{(1)})$ and $(C_{\varphi WB}, C_{\varphi u})$. We analyze the joint fits from each DIS experiment, as well as the electroweak-pole-observable (EWPO) fits adapted from Ref.~\cite{Ellis:2020unq}
From the $2d$ projections in Fig.~\ref{fig:ewpo-comparison2}, we can see that the potential LHeC probes are stronger than those of the joint electroweak and LHC fit and that the FCC-eh bounds are even stronger. In particular, the joint electroweak and LHC fit exhibits strong correlations between parameters that result in elongated ellipses in several of the $2d$ projections that we consider, as illustrated by the pair $(C_{\varphi WB}, C_{\varphi \ell}^{(1)})$. The combinations of future LHeC and FCC-eh runs do not show these correlations, and can remove these (approximate) degeneracies in the joint electroweak and LHC fit.
We note that the EIC probes are far weaker than those obtained from the other fits, and do not contribute significantly to probing the $ffV$ parameter space.

\section{Conclusions}
\label{s:conclusions}

This study investigates the potential of the LHeC, FCC-eh, and EIC in exploring BSM physics within the framework of the SMEFT. Building upon previous research, we focus on key observables: the NC DIS cross section at the LHeC and the FCC-eh, and PV asymmetries at the EIC. By considering SMEFT semi-leptonic four-fermion operators and $ffV$ vertex corrections, we work first with a 7-dimensional and then with a 17-dimensional Wilson coefficient parameter space. 

Our $7d$ fits reveal that at the EIC can probe UV scales up to 700 GeV. For the LHeC, the range reaches 14 TeV, while at the FCC-eh, the probed scales span from 2 TeV to 18 TeV.
We note that no single-run scenario at any of these experiments provides an ideal probe for the complete SMEFT parameter space. 
% Varying polarization and lepton species is crucial for effective BSM studies. 
Our study demonstrates that future precision DIS measurements can effectively alleviate degeneracies observed in precision electroweak fits based on $Z$-pole observables. The constraints obtained from the LHeC and FCC-eh experiments are generally stronger compared to the combined fits using $Z$-pole and LHC data. Overall, our results underscore the considerable potential of future DIS studies in exploring BSM physics.

%%%%%%%%%%%%%%%%%%%%%%%%%%%%%%%%%%%%%%%%%%%%%%%%%%%%%%%%%%%%%%%%%%%%%%%%%%%%%%%%%%%%%%%%%%%%
\vspace{0.2cm}
%\begin{acknowledgments}
\emph{Acknowledgments.}
 We thank D. Britzger for suggesting to include an analysis of the FCC-eh capabilities.
 C. B. and R. B. are supported by the DOE contract DE-AC02-06CH11357. K. S. is supported by the DOE grant DE- FG02-91ER40684. This research was supported in part through the computational resources and staff contributions provided for the Quest high performance computing facility at Northwestern University which is jointly supported by the Office of the Provost, the Office for Research, and Northwestern University Information Technology.
%\end{acknowledgments}

%%%%%%%%%%%%%%%%%%%%%%%%%%%%%%%%%%%%%%%%%%%%%%%%%%%%%%%%%%%%
\bibliographystyle{JHEP}
\bibliography{DIS23proceedings.bib}

\providecommand{\href}[2]{#2}\begingroup\raggedright\begin{thebibliography}{10}

\bibitem{Brivio:2017vri}
I.~Brivio and M.~Trott, \emph{{The Standard Model as an Effective Field
  Theory}}, \href{http://dx.doi.org/10.1016/j.physrep.2018.11.002}{\emph{Phys.
  Rept.} {\bf 793} (2019) 1--98}, [\href{http://arxiv.org/abs/1706.08945}{{\tt
  1706.08945}}].

\bibitem{Bissolotti:2023vdw}
C.~Bissolotti, R.~Boughezal and K.~Simsek, \emph{{SMEFT probes in future
  precision DIS experiments}},  \href{http://arxiv.org/abs/2306.05564}{{\tt
  2306.05564}}.

\bibitem{Britzger:2020kgg}
D.~Britzger, M.~Klein and H.~Spiesberger, \emph{{Electroweak physics in
  inclusive deep inelastic scattering at the LHeC}},
  \href{http://dx.doi.org/10.1140/epjc/s10052-020-8367-y}{\emph{Eur. Phys. J.
  C} {\bf 80} (2020) 831}, [\href{http://arxiv.org/abs/2007.11799}{{\tt
  2007.11799}}].

\bibitem{Boughezal:2022pmb}
R.~Boughezal, A.~Emmert, T.~Kutz, S.~Mantry, M.~Nycz, F.~Petriello et~al.,
  \emph{{Neutral-current electroweak physics and SMEFT studies at the EIC}},
  \href{http://dx.doi.org/10.1103/PhysRevD.106.016006}{\emph{Phys. Rev. D} {\bf
  106} (2022) 016006}, [\href{http://arxiv.org/abs/2204.07557}{{\tt
  2204.07557}}].

\bibitem{Boughezal:2020uwq}
R.~Boughezal, F.~Petriello and D.~Wiegand, \emph{{Removing flat directions in
  standard model EFT fits: How polarized electron-ion collider data can
  complement the LHC}},
  \href{http://dx.doi.org/10.1103/PhysRevD.101.116002}{\emph{Phys. Rev. D} {\bf
  101} (2020) 116002}, [\href{http://arxiv.org/abs/2004.00748}{{\tt
  2004.00748}}].

\bibitem{Boughezal:2021kla}
R.~Boughezal, F.~Petriello and D.~Wiegand, \emph{{Disentangling Standard Model
  EFT operators with future low-energy parity-violating electron scattering
  experiments}},
  \href{http://dx.doi.org/10.1103/PhysRevD.104.016005}{\emph{Phys. Rev. D} {\bf
  104} (2021) 016005}, [\href{http://arxiv.org/abs/2104.03979}{{\tt
  2104.03979}}].

\bibitem{Boughezal:2023ooo}
R.~Boughezal, D.~de~Florian, F.~Petriello and W.~Vogelsang, \emph{{Transverse
  spin asymmetries at the EIC as a probe of anomalous electric and magnetic
  dipole moments}},
  \href{http://dx.doi.org/10.1103/PhysRevD.107.075028}{\emph{Phys. Rev. D} {\bf
  107} (2023) 075028}, [\href{http://arxiv.org/abs/2301.02304}{{\tt
  2301.02304}}].

\bibitem{Grzadkowski:2010es}
B.~Grzadkowski, M.~Iskrzynski, M.~Misiak and J.~Rosiek, \emph{{Dimension-Six
  Terms in the Standard Model Lagrangian}},
  \href{http://dx.doi.org/10.1007/JHEP10(2010)085}{\emph{JHEP} {\bf 10} (2010)
  085}, [\href{http://arxiv.org/abs/1008.4884}{{\tt 1008.4884}}].

\bibitem{deFlorian:2012wk}
D.~de~Florian and Y.~Rotstein~Habarnau, \emph{{Polarized semi-inclusive
  electroweak structure functions at next-to-leading-order}},
  \href{http://dx.doi.org/10.1140/epjc/s10052-013-2356-3}{\emph{Eur. Phys. J.
  C} {\bf 73} (2013) 2356}, [\href{http://arxiv.org/abs/1210.7203}{{\tt
  1210.7203}}].

\bibitem{Altarelli:1979kv}
G.~Altarelli, R.~K. Ellis, G.~Martinelli and S.-Y. Pi, \emph{{Processes
  Involving Fragmentation Functions Beyond the Leading Order in QCD}},
  \href{http://dx.doi.org/10.1016/0550-3213(79)90062-2}{\emph{Nucl. Phys. B}
  {\bf 160} (1979) 301--329}.

\bibitem{Vogelsang:1990ug}
W.~Vogelsang, \emph{{The Gluonic contribution to g(1)-p (x,Q**2) in the parton
  model}}, \href{http://dx.doi.org/10.1007/BF01474080}{\emph{Z. Phys. C} {\bf
  50} (1991) 275--284}.

\bibitem{Altarelli:1979ub}
G.~Altarelli, R.~K. Ellis and G.~Martinelli, \emph{{Large Perturbative
  Corrections to the Drell-Yan Process in QCD}},
  \href{http://dx.doi.org/10.1016/0550-3213(79)90116-0}{\emph{Nucl. Phys. B}
  {\bf 157} (1979) 461--497}.

\bibitem{deFlorian:1994wp}
D.~de~Florian and R.~Sassot, \emph{{O (alpha-s) spin dependent weak structure
  functions}}, \href{http://dx.doi.org/10.1103/PhysRevD.51.6052}{\emph{Phys.
  Rev. D} {\bf 51} (1995) 6052--6058},
  [\href{http://arxiv.org/abs/hep-ph/9412255}{{\tt hep-ph/9412255}}].

\bibitem{r:KleinData}
``http://hep.ph.liv.ac.uk/\~{}mklein/lhecdata/.''

\bibitem{r:KleinPaper}
M.~Klein and V.~Radescu Tech. Rep. CERN-LHeC-Note-2013-002PHY, 2013.

\bibitem{Britzger:2022abi}
D.~Britzger, M.~Klein and H.~Spiesberger, \emph{{Precision electroweak
  measurements at the LHeC and the FCC-eh}},
  \href{http://dx.doi.org/10.22323/1.398.0485}{\emph{PoS} {\bf EPS-HEP2021}
  (2022) 485}, [\href{http://arxiv.org/abs/2203.06237}{{\tt 2203.06237}}].

\bibitem{Alte:2018xgc}
S.~Alte, M.~K\"onig and W.~Shepherd, \emph{{Consistent Searches for SMEFT
  Effects in Non-Resonant Dilepton Events}},
  \href{http://dx.doi.org/10.1007/JHEP07(2019)144}{\emph{JHEP} {\bf 07} (2019)
  144}, [\href{http://arxiv.org/abs/1812.07575}{{\tt 1812.07575}}].

\bibitem{Dawson:2019clf}
S.~Dawson and P.~P. Giardino, \emph{{Electroweak and QCD corrections to $Z$ and
  $W$ pole observables in the standard model EFT}},
  \href{http://dx.doi.org/10.1103/PhysRevD.101.013001}{\emph{Phys. Rev. D} {\bf
  101} (2020) 013001}, [\href{http://arxiv.org/abs/1909.02000}{{\tt
  1909.02000}}].

\bibitem{Ellis:2020unq}
J.~Ellis, M.~Madigan, K.~Mimasu, V.~Sanz and T.~You, \emph{{Top, Higgs, Diboson
  and Electroweak Fit to the Standard Model Effective Field Theory}},
  \href{http://dx.doi.org/10.1007/JHEP04(2021)279}{\emph{JHEP} {\bf 04} (2021)
  279}, [\href{http://arxiv.org/abs/2012.02779}{{\tt 2012.02779}}].

\end{thebibliography}\endgroup
%%%%%%%%%%%%%%%%%%%%%%%%%%%%%%%%%%%%%%%%%%%%%%%%%%%%%%%%%%%%
\end{document}